\documentclass[aip
 amsmath,amssymb,
 reprint,%
]{revtex4-2}

\usepackage{graphicx}
\usepackage{dcolumn}
\usepackage{bm}
\usepackage{xcolor}
\usepackage[utf8]{inputenc}
\usepackage[T1]{fontenc}
\usepackage{mathptmx}
\usepackage{amsmath}
\usepackage{etoolbox}
\usepackage{float}

\makeatletter
\def\@email#1#2{%
 \endgroup
 \patchcmd{\titleblock@produce}
  {\frontmatter@RRAPformat}
  {\frontmatter@RRAPformat{\produce@RRAP{*#1\href{mailto:#2}{#2}}}\frontmatter@RRAPformat}
  {}{}
}%
\makeatother
\begin{document}

\preprint{AIP/123-QED}

\title{Super resonance: \\
Breaking the bandwidth limit of resonant modes and its application to flow control}

\author{Adam R. Harris}
\affiliation{Ann and H.J. Smead Department of Aerospace Engineering Sciences, University of Colorado Boulder, Boulder, Colorado 80303, USA}
\affiliation{Materials Science and Engineering Program, University of Colorado Boulder, Boulder, Colorado 80303, USA}
\author{Armin Kianfar}
\affiliation{Ann and H.J. Smead Department of Aerospace Engineering Sciences, University of Colorado Boulder, Boulder, Colorado 80303, USA}
\author{David Roca}
\affiliation{Centre Internacional de M\`{e}todes Num\`{e}rics en Enginyeria (CIMNE), Barcelona 08034, Spain}
\affiliation{Universitat Polit\`{e}cnica de Catalunya, ESEIAAT Campus Terrassa UPC, Terrassa 08222, Spain}
\author{Daniel Yago}
\affiliation{Centre Internacional de M\`{e}todes Num\`{e}rics en Enginyeria (CIMNE), Barcelona 08034, Spain}
\affiliation{Universitat Polit\`{e}cnica de Catalunya, ESEIAAT Campus Terrassa UPC, Terrassa 08222, Spain}
 \author{Christoph Brehm}
\affiliation{Department of Aerospace Engineering, University of Maryland, College Park, Maryland 20740, USA}
\author{Mahmoud I. Hussein}%
\email{mih@colorado.edu}
\affiliation{Ann and H.J. Smead Department of Aerospace Engineering Sciences, University of Colorado Boulder, Boulder, Colorado 80303, USA}
\affiliation{Department of Physics, University of Colorado Boulder, Boulder, Colorado 80302, USA}
\affiliation{Materials Science and Engineering Program, University of Colorado Boulder, Boulder, Colorado 80303, USA}


\begin{abstract}

We report the discovery of super resonance—a new regime of resonant behavior in which a mode’s out-of-phase response persists far beyond its classical bandwidth. This effect emerges from a coiled phononic structure composed of a locally resonant elastic metamaterial and architected to support multiple internal energy pathways. These pathways converge at a single structural location, enabling extended modal dominance and significantly broadening the frequency range over which a resonant phase is sustained. We demonstrate by direct numerical simulations the implications of this mechanism in the context of flow instability control, where current approaches are inherently constrained by the characteristically narrow spectral bandwidth of conventional resonances. Using a super-resonant phononic subsurface structure interfacing with a channel flow, we show passive simultaneous suppression of four unstable flow perturbations across a frequency range more than five times wider than that is achievable with a standard resonance in an equivalent uncoiled structure. By enabling broadband, passive control of flow instabilities, super resonance overcomes a longstanding limitation in laminar flow control strategies. More broadly, it introduces a powerful new tool for phase-engineered wave–matter interaction. The ability to preserve out-of-phase modal response across wide spectral ranges establishes a fundamental advance in the physics of resonance, with far-reaching implications for suppressing fully developed turbulent flows and beyond.

\end{abstract}
\vspace*{1ex}

\maketitle



\noindent \textbf{INTRODUCTION}

The harmonic oscillator is one of the most fundamental constructs in physics, serving as a building block across classical mechanics, quantum theory, solid-state physics, and wave dynamics~\cite{Lagrange1788,born1924quantum,dirac1927quantum,feynman1965quantum,ashcroft1976solid}. Its resonance behavior—marked by usually a sharply peaked amplitude response and an associated narrow out-of-phase spectral region—underpins our understanding of modal systems, from phonons and photons to acoustic and electromagnetic cavities.~In both isolated and coupled forms, harmonic oscillator modes are intrinsically limited in bandwidth; for coupled systems, the phase response is tightly overridden by the presence of nearby resonances or anti-resonances.~This long-standing narrowband constraint has limited the scope of resonance-based control strategies across physics and engineering. In this work, we introduce \textit{super resonance}: a new regime of modal behavior in which a structural mode sustains an out-of-phase response that persists far beyond its classical resonance bandwidth, despite the nominal presence of adjacent resonant peaks or anti-peaks.~This behavior is achieved through a uniquely architected coiled phononic structure featuring multiple internal elastic energy pathways that converge to an effectively single spatial location.~This configuration enables extended modal phase persistence, laying the foundation for broadband control in systems traditionally constrained by narrowband resonance dynamics—including the passive suppression of multiple unstable flow perturbations, which is the focus of this investigation.\\

\noindent \textbf{\textit{Flow control}}\\
\indent Fluid-structure interaction is a dynamical process that is central to managing skin-friction drag in vehicles operating in air, water, or on land, as well as in applications like turbomachinery~\cite{Gad2003,Tiainen2017}.~The magnitude of skin-friction drag, particularly for streamlined bodies, is a key factor in determining fuel efficiency—lower drag results in higher efficiency.~A significant reduction in skin-friction drag can be achieved by delaying the transition from laminar to turbulent flow.~Minimizing or delaying this effect is therefore a primary objective in flow control.~This fundamental flow transformation is tied directly with the behavior of unstable flow disturbances, or perturbations, that occur naturally in a flow~\cite{morkovin1969many}.~Among the most studied are Tollmien–Schlichting (TS) waves~\cite{schlichting2016boundary}. The linear modal nature of these waves provides an opportunity to apply phased control with a prescribed external stimulus to inhibit their growth by wave superposition~\cite{milling1981tollmien}.~Numerous studies have explored the implementation of this active control approach using various techniques~\cite{liepmann1982control,liepmann1982active,thomas1983control,Joslin_1995,Grundmann_2008,Amitay_2016}.~However, these methods require energy input and intricate sensing and actuation systems for feedback control~\cite{hu1994feedback,bewley1998optimal}. In the absence of closed-loop feedback, phase locking to a specific unstable wave becomes necessary to time the intervention correctly~\cite{Amitay_2016}, which restricts the ability to manage multiple or spontaneously occurring instabilities.\\ 
\indent To overcome these limitations, the concept of phononic subsurfaces (PSubs) has been introduced as a passive intervention approach capable of precise and responsive control of flow instabilities~\cite{Hussein_2015}.~A PSub comprises a finite phononic structure that is placed beneath the fluid-structure interface.~Given its finite size, it represents a truncated~\cite{davis2011analysis,al2023theory,rosa2023material} phononic material~\cite{deymier2013acoustic,hussein2014dynamics,Phani_2017,Jin_2021}, such as a phononic crystal that exhibits Bragg scattering~\cite{Hussein_2015,Barnes_2021,kianfar2023phononicNJP,michelis2023attenuation,schmidt2025perturbation}, or a locally resonant elastic metamaterial that exhibits resonance hybridizations~\cite{kianfar2023phononicNJP,kianfar2023local}.~The unit-cell and finite-structure characteristics of a PSub may be designed to passively force incident flow instability waves to encounter out-of-phase conditions. This causes a decrease, near the surface, in the production rate of the kinetic energy of the continuously incoming flow perturbation waves, and subsequently their attenuation across the region covered by the PSub.~The outcome is a local reduction in skin-friction drag where the PSub unit is installed~\cite{kianfar2023local}.~Recent developments extended the concept to downstream control, by either adopting a multi-input, multi-output configuration~\cite{willey2023multi}, employing a lattice of PSubs~\cite{hussein2025scatterless}, or using PSubs tuned to intermediate phase values~\cite{klauss2025control}.~The tuning of a single PSub structure depends on knowledge of the frequencies and wavenumbers of the relevant flow instabilities, including incidents involving multiple coexisting modes. \\
\indent While this passive phonon-enabled approach has shown great promise, in areas ranging from transition delay~\cite{willey2023multi,hussein2025scatterless} to control of unsteady flows~\cite{wiberg2025one}, and recently extending to disturbance attenuation in laminar hypersonic flows~\cite{klauss2025control}, its effectiveness remains limited by the characteristically narrowband nature of the modal response in the interfacing structure.~This limitation also applies to other conventional resonators such as Helmholtz resonators~\cite{michelis2023interaction}, and other wall-based flow control approaches suffer the same constraint~\cite{zhao2022review}.~Extensive investigations have been done on various PSub configurations, but due to this constraint are all confined to spectrally narrow phase response—whether, for example, due to a truncation resonance~\footnote{A band-gap truncation resonance may exhibit a modestly broader out-of-phase bandwidth than a standard structural resonance—particularly when the band gap is wide~\cite{kianfar2023phononicNJP}.~Yet it remains fundamentally a conventional resonance subject to the same narrowband limitations described above.} located within a band gap~\cite{Hussein_2015,Barnes_2021,michelis2023attenuation} or a subwavelength structural resonance influenced by the presence of local resonators~\cite{kianfar2023phononicNJP, kianfar2023local}.~This narrowband restriction reduces the applicability of PSubs to realistic settings, such as aircraft flight, where flow instabilities emerge over a relatively broad frequency range~\footnote{These instability bands depend on the Reynolds number $Re$ and can be predicted \textit{a priori} through linear stability analysis~\cite{schubauer1947laminar,mack1984boundary,Barnes_2021}}.~In this paper, by introducing super resonance, we present a new PSub concept that facilitates a significant broadening of the frequency range of operation.\\

\noindent \textbf{\textit{Super resonance}}\\
\indent Numerous studies across the physical sciences have investigated different forms of enhanced resonant behavior, often highlighting superior properties.~Examples include acoustic super-scattering from elastic shells~\cite{tolstoy1986superresonant}, strong plasmonic field enhancement and lasing~\cite{gordon2007design}, high-Q supercavity modes enabled by bound states in the continuum in subwavelength dielectric resonators~\cite{rybin2017high}, and intensified Mie resonances in dielectric microspheres~\cite{wang2019high}.~In each case, the response is intensified through distinct means—boundary-induced coherence, material gain, non-radiating interference effects, or structural symmetry, respectively.~However, the underlying mechanisms remain rooted in conventional resonant behavior: a narrowband spectral peak that fundamentally constrains modal reach across the frequency domain.~In contrast, the current work targets broadening the phase bandwidth of a target resonance beyond its natural limits. The core premise is to intervene in key energy pathways that govern the mechanistic formation of the resonance as it is observed at the response location.\\
\indent In any resonant system, a peak response arises due to an excitation at a specific spatial location (or domain), and, similarly, there is a distinct spatial location (or domain) where the response may be probed.~For a grounded simple harmonic oscillator, there is only a single point of excitation and a single response point, creating a resonant peak with its characteristic narrow bandwidth~\footnote{The width of a resonant peak may be broadened by adding dissipation, but at the expense of the response amplitude}.~Such characteristic narrow band also holds for modes in a coupled oscillator system.~Here we consider a multi-degree-of-freedom system composed of coupled oscillators.~But we reconfigure the spatial topology to allow multiple pathways for the energy within the structure to converge towards an effectively single response point in space, selected to coincide with the excitation point. In this manner, the modal response of the system exhibits an effectively reconstructed mode with a significantly widened bandwidth, exceeding the fundamental bandwidth associated with a conventional resonant mode. \\
\indent For application to the flow stabilization problem, the concept is realized in the form of a PSub architecture composed of a coiled beam structure, configured such that more than one location along the structure interfaces with \textit{a single point} in the flow. The coiling is done by a series of full $180^{\circ}$ turns, incorporating rotational locking at the turning junctions$-$an approach that fully preserves the phonon band structure compared to the uncoiled configuration~\cite{willey2022coiled}. By following this design strategy, we observe remarkable broadening of a performance metric that governs the strength and spectral extent of intervention with the flow~\cite{Hussein_2015,kianfar2023phononicNJP,kianfar2023local}.\\
\begin{figure*} [t]
\centering
\includegraphics[width=1\textwidth]{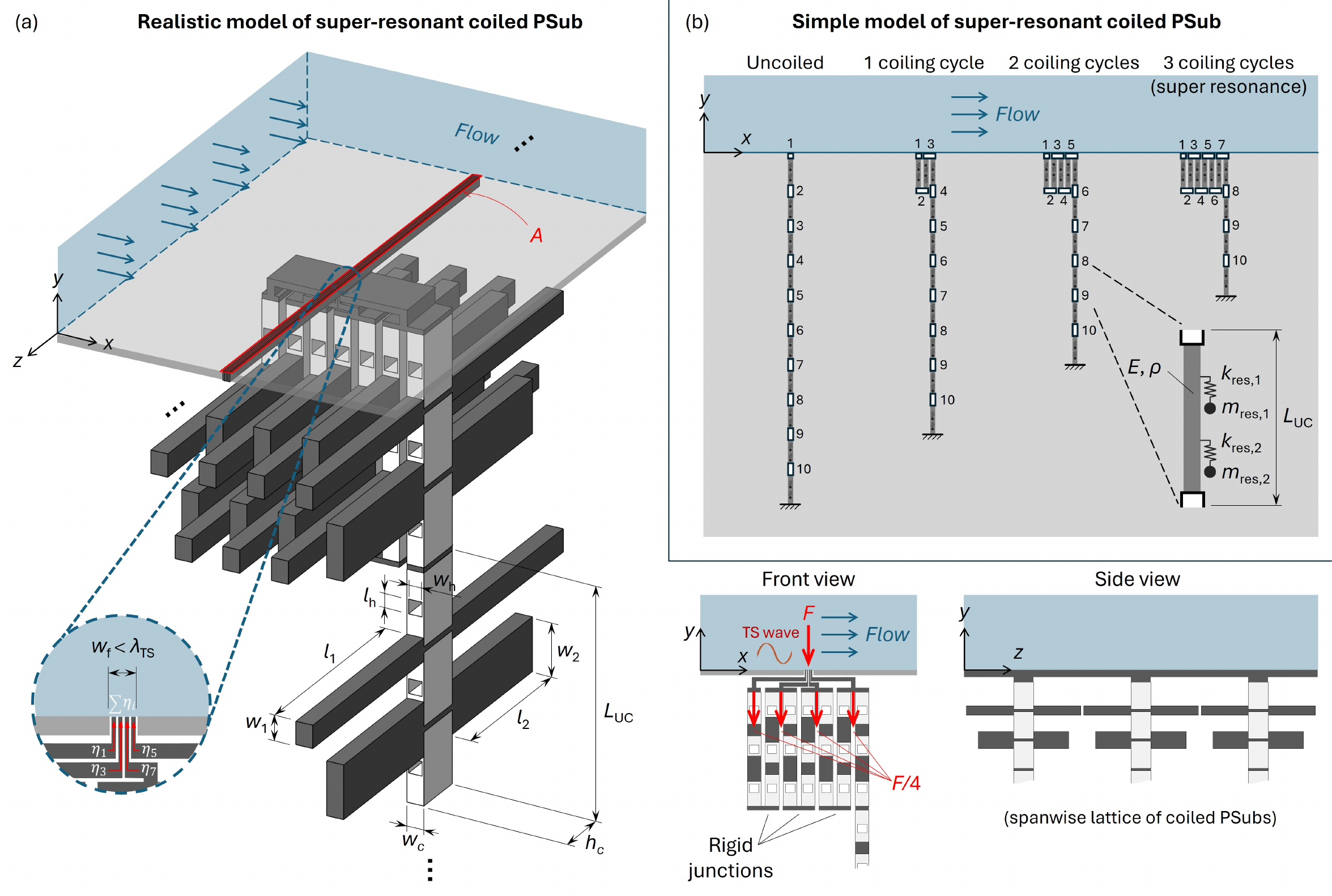}
\caption{\textbf{Super-resonant coiled PSub design:}~A schematic of the concept of a coiled phononic subsurface with multiple structural connectivity. The physical model, shown in (a), features a central voided beam that is rotationally locked at the turning junctions, leading to a full preservation of the phonon band structure~\cite{willey2022coiled}.~The beam ``trunk" has an array of branching resonators in the form of small cantilevered beams$-$this configuration in its entirety is amenable to fabrication.~The dimensions of the PSub unit cell are as follows based on the parameters shown in (a): $L_{\text{UC}} = 20$ mm, $w_{\text{c}} = 1.44$ mm, $h_{\text{c}} = 4.63$ mm, $l_{\text{h}} = 1.17$ mm, $w_{\text{h}} = 1.12$ mm, $l_1 = 18.23$ mm, $w_1 = 2.06$ mm, $l_2 = 12.93$ mm, $ w_2 = 4.68$ mm, $w_{\text{f}} = 0.4$ mm.~The central beam component is composed of ABS polymer with density $\rho_{\text{ABS}}=1200$ kg/m$^3$ and Young's modulus $E_{\text{ABS}}=1$ GPa.~The resonant branches are made from aluminium with density $\rho_{\text{Al}} = 2700$ kg/m$^3$ and Young's modulus $E_{\text{Al}} = 68.8$ GPa.~In our computational investigation, this 3D structure is modeled in the form of a 1D rod, as shown in (b), with effective Young's modulus $E_{\text{eff}} = 0.2$ GPa and density $\rho_{\text{eff}} = 450$ kg/m$^3$, containing an array of spring-mass resonators attached based on the following parameters: $L_{\text{UC}} = 20$ mm, $k_{\text{res},1} = 2.6 \times 10^{10}$ N/m, $k_{\text{res},2} = 7.11 \times 10^{11}$ N/m, $m_{\text{res},1} = 1.62$ g, $m_{\text{res},2} = 4.2$ g.~Material  damping is introduced to the rod component in the form of viscous proportional damping with constants $q_1=0$ and $q_2=6\times 10^{-8}$~\cite{kianfar2023phononicNJP}.~Given that our interest is in only the longitudinal motion, the rod model provides a good approximation of the coiled PSub response, owing to the rotational locking applied at the junctions.~The selected simple model parameters deviate less than 2\% from the actual unit-cell equivalent configuration.~With one coiling cycle, we have two structural points that are simultaneously interfacing with the flow, labeled Junctions 1 and 3. With two coiling cycles, Junctions 1, 3, and 5 are in contact with the flow, and so on in an odd-number progression.~We find that with three coiling cycles, comprising Junctions 1, 3, 5, 7, super-resonant behavior emerges (see Fig.~\ref{fig:F2}).~While the coiled PSub design is confined to a specific finite dimension along the spanwise direction, its response can still be considered one-dimensional when normalized by the surface area interacting with the flow, marked in red and denoted by $A$.~The excitation of the structure is marked by $F$ and the corresponding input into the flow is marked by $\Sigma\eta_{i}$.~In this work, we consider only a single coiled PSub covering the full span of the channel.~However, application to longer spanwise distances (e.g., for wider channels or boundary-layer flows) may be realized by installation of a periodic arrangement of the coiled PSub along that direction (side view).~Additional rows may also be added along the streamwise direction to form a PSub lattice setup~\cite{hussein2025scatterless}.}
\label{fig:F1}
\end{figure*}
\begin{figure*} [t!]
\centering
\includegraphics[width=0.85\textwidth]{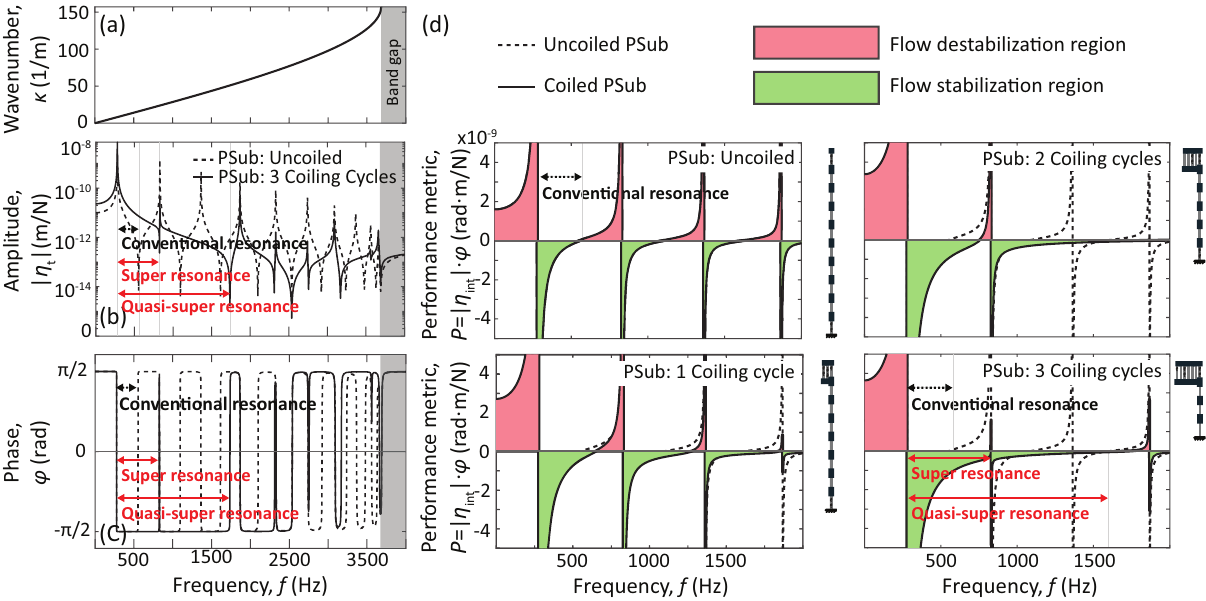}
\caption{\textbf{Evidence of super resonance: Coiled PSub dispersion and vibration characteristics:}~(a) Dispersion, (b-c) harmonic frequency response, and (d) performance metric for the coiled PSub with multiple structural connectivity. The dispersion is unchanged with the number of coiling cycles since the design of the unit cell is the same in all cases and band structure preservation is maintained due to the rotational lockings~\cite{willey2022coiled}.~The amplitude and phase frequency dependency for the uncoiled (0 coils) and 3-cycle coiled configurations are shown in (b-c).~Positive (red) and negative (green) regions of $P$ represent destabilization and stabilization, respectively, upon employment of the PSub for flow control.~Zero coiling cycles represent a PSub with 1 point-of-contact with the flow, while 3 coiling cycles represents the same PSub but with 4 points-of-contact with the flow, as demonstraed in Fig.~\ref{fig:F1}.~As the number of coils is increased from 1 to 3, the destabilization and stabilization regions across the frequency domain are narrowed and broadened, respectively. With 3 coiling cycles, super resonance emerges yielding a perfectly contiguous stabilization band that is nearly two times larger than that of the uncoiled reference case.~A quasi-super resonance regime also emerges, one that is effectively contiguous and nearly four times larger than that of the reference configuration.}
\label{fig:F2}
\end{figure*}
\begin{flushleft}
\textbf{RESULTS AND DISCUSSION}
\end{flushleft}

\noindent \textbf{Unstable channel flow and coiled PSub} \\
\indent To establish the design conditions for the super-resonant PSub, we begin by describing the flow problem. Governed by the three-dimensional (3D) Naiver-Stokes equations, our analysis involves a series of direct numerical simulations (DNS) for incompressible channel flows. The velocity vector solution is expressed as ${\bf u}(x,y,z,t)=(u,v,w)$ with components in the streamwise $x$, wall-normal $y$, and the spanwise $z$ directions, respectively, where $t$ denotes time.~We run the DNS for a Reynolds number of $Re=\rho_\mathrm{f} U_{\rm c}\delta/\mu_\mathrm{f}=7500$ based on a centerline velocity $U_{\rm c}=11.484$ $\mathrm{m/s}$ and a half-height of the channel $\delta=6.53\times10^{-4}$ $\mathrm{m}$. Liquid water is considered with a density of $\rho_\mathrm{f}=1000$  $\mathrm{kg/m^3}$ and dynamic viscosity of $\mu_\mathrm{f}=1\times10^{-3}$ $\mathrm{kg/ms}$~\footnote{Future work may follow the same methodology for air or other fluids.}. 
All subsequent quantities in this paper, unless mentioned explicitly, are normalized by the channel's centerline velocity $U_{\rm c}$ and half-height $\delta$, representing the outer velocity and length scales, respectively. Computational details on our DNS scheme are presented in the Appendix.\\
\indent The channel size is $0 \leq x \leq 30$, $0 \leq y \leq 2$, and $0 \leq z \leq 2 \pi$ for the streamwise, wall-normal, and spanwise directions, respectively.~Flow simulations are initialized by superimposing a fully developed Poiseuille flow with unstable spatially developing TS modes obtained from linear stability analysis governed by the Orr-Sommerfeld equation \cite{Orr1907,Sommerfeld1909} and solved for the same $Re$. 
We select four unstable eigensolutions within the non-dimensional frequency range
$0.205 \le \omega_\mathrm{TS} \le 0.283$, as frequencies outside of this range represent stable TS modes. The four spatial eigensolutions have complex wavenumbers $\alpha_{1}=0.9014-\mathrm{i}0.002576$, $\alpha_{2}=0.9508-\mathrm{i}0.005579$, $\alpha_{3}=1.0004-\mathrm{i}0.006171$, $\alpha_{4}=1.0502-\mathrm{i}0.004131$ and real non-dimensional frequencies $\omega_\mathrm{TS,1}=0.214$, $\omega_\mathrm{TS,2}=0.232$, $\omega_\mathrm{TS,3}=0.250$, $\omega_\mathrm{TS,4}=0.268$, respectively.~Following dimensional analysis (based on $U_{\text{c}}$ and $\delta$), the frequencies of these TS modes are $\Omega_\mathrm{TS}={\omega_\mathrm{TS}}{U_{\rm c}}/{2\pi}{\delta}$ where $\Omega_\mathrm{TS,1} = 600$ $\mathrm{Hz}$, $\Omega_\mathrm{TS,2} = 650$ $\mathrm{Hz}$, $\Omega_\mathrm{TS,3} = 700$ $\mathrm{Hz}$, and $\Omega_\mathrm{TS,4} = 750$ $\mathrm{Hz}$. To ensure outgoing waves on the other side of the channel, the disturbances are smoothly brought to zero by attaching a non-reflective buffer region at the outlet~\cite{Dana91,saiki1993spatial,kucala2014spatial}.~Periodic boundary conditions are applied in the spanwise direction.~At the top and bottom walls, no-slip/no-penetration boundary conditions are applied, except within the control region from $x_\mathrm{s}$ to $x_\mathrm{e}$ in the streamwise direction where the rigid wall is replaced by a coiled PSub at the bottom wall.~Within the control region, the fluid-structure coupling is enforced by means of transpiration boundary conditions~\cite{Lighthill_1958,Sankar_1981,Hussein_2015,kianfar2023phononicNJP} (see Appendix).~These boundary conditions are valid if the PSub motion is only in the wall-normal direction and $\eta_{\text{int}} \ll \delta$ where $\eta_{\text{int}}$ is the total wall-normal displacement of the flow-interfacing junctions of the PSub (see below).~Hence, throughout the DNS, the roughness Reynolds number is monitored and maintained below 25.~This ensures that the flow remains hydraulically smooth~\cite{morkovin1990roughness}, justifying the use of static grid points in DNS.~A variety of flow quantities are calculated for an indepth analysis on the effects of the PSubs; these are defined and described in the Appendix.\\
\indent In its nominal uncoiled form, the PSub considered comprises a finite linear elastic metamaterial consisting of 10 unit cells, where each is formed from three central voided beam-like segments with two distinct branching cantilever beams in-between, serving as local resonators; see illustration with geometric details in Fig.~\ref{fig:F1}.~To make extensive coupled fluid-structure simulations feasible, this phononic structure is modeled as an elastic rod with two mass-spring resonators attached per unit cell, similar to the configuration of Ref.~\cite{kianfar2023phononicNJP}.~The underlying reduce-order-modeling is facilitated by a combined homogenization and machine learning-algorithm~\cite{sal2023optimal,yago2024machine}; see Appendix for details.~The PSub is free to deform at the edge interfacing with the flow (top) as well as all other intermediate locations throughout its structure, but is fixed at the other end (bottom). The PSub elastic domain is allowed to deform in complete independence from the adjacent rigid wall, thus the flow ``experiences" a deformation across a given fluid-PSub interface junction~\cite{Hussein_2015,kianfar2023phononicNJP}. The length of the unit cell along the wall-normal direction is $L_\mathrm{UC}= 2$ $\mathrm{cm}$ (i.e., total uncoiled PSub length is $20$ $\mathrm{cm}$).~The resonator frequencies are set to $\Omega_\mathrm{res,1}=5000$ $\mathrm{Hz}$ and $\Omega_\mathrm{res,2}=20,000$ $\mathrm{Hz}$ by tuning their point masses to be three and five times heavier than the total mass of the unit-cell base, i.e., $m_\mathrm{res,1}=3\times \rho L_\mathrm{UC}$ and $m_\mathrm{res,2}=5\times \rho L_\mathrm{UC}$, where $\rho$ is the base material effective density.~Hence, the stiffnesses of the resonator springs are $k_\mathrm{res,n}=m_\mathrm{res,n}(2\pi f_\mathrm{res,n})^2$, where $n=1,2$.\\
~\indent The PSub coiling protocol is done as follows. A full $180^{\circ}$ turn is applied to each unit cell, starting with the second one, and up to twice the number of coiling cycles.~Thus, with $n$ coiling cycles, $2n$ full turns are applied at the intersections of the first $2n+1$ unit cells.~In order to guarantee that the phonon band structure is preserved, rotationally rigid junctions are placed at these intersections to lock the rotation degree of freedom~\cite{willey2022coiled}.~On the top side, these junctions converge to a $n+1$ points in contact with the flow, with the ability to interact independently with it. At these points, the flow represents the excitation source from the view point of the coiled PSub.~Since these points are confined within a small control region along the streamwise direction compared to the perturbation wavelength(s), from the flow point of view they are effectively perceived as a single response point.~This means that the rotationally rigid junctions transmit and distribute the total force equally from the flow to the coiled PSub, and the coiled PSub’s response is the cumulative contribution of all the points in contact with the flow.In this manner, the fluid-structure interaction takes a multiple input, multiple output form for the PSub and a single input, single output form for the flow.~See Fig.~\ref{fig:F1} for a schematic and geometric details.\\
\begin{figure*} [t!]
\centering
\includegraphics[width=0.9\textwidth]{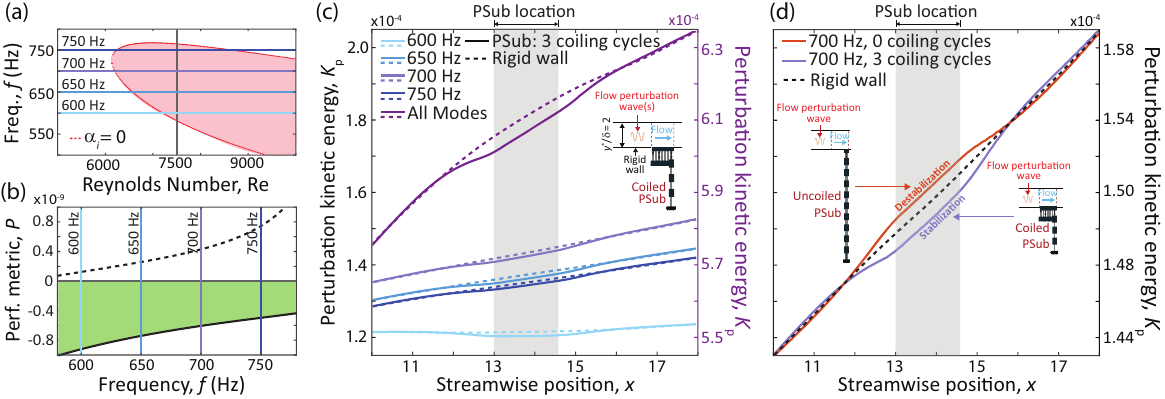}
\centering
\caption{\textbf{Super-resonant coiled PSub performance by DNS:}~(a) Stability map in the frequency-Reynolds number domain obtained by solving the Orr-Sommerfeld equation~\cite{Orr1907,Sommerfeld1909} for the channel.~(b) Zoom-in of the performance metric plot for the super-resonant 3-cycle coiled PSub, with frequencies of the four unstable TS waves marked.~(c)~Time-averaged perturbation kinetic energy in the flow as a function of streamwise position in the channel.~The results of five separate simulations are shown: four in which a distinct TS instability wave is controlled (left axis), and one where all four TS instability waves are controlled simultaneously (right axis).~In all cases, the 3-cycle coiled PSub successfully causes local stabilization in the flow, with stabilization strength consistent with the corresponding value of the performance metric.~The all-modes case demonstrates broadband stabilization spanning the entire unstable region for the given value of $Re$ [shown in (a)] with an intensity that matches the ''additive" effect of the $P$ values over the four instability frequencies [shown in (b)].~(d) $K_{\text{p}}$ plot for the 700-Hz instability case showing contrast in performance when the 3-cycle coiled Psub is installed versus using the nominal uncoiled PSub.~The results confirm the transformation from destabilization to stabilization due to the coiling.}
\label{fig:F3}
\end{figure*}

\noindent \textbf{Coiled PSub response characteristics} \\
\indent The results shown in Fig.~\ref{fig:F2} are obtained by FE analysis, separate from the subsequent fluid-structure simulations we discuss later.~The coiled PSub turning junctions at which the rotational locking is applied are denoted Junctions $i=\{1,2,3$...$,10\}$ in Fig.~\ref{fig:F1}b.~The dispersion curves for longitudinal motion for the resonant elastic PSub as a wave propagation medium as well as the frequency response of its finite structural version are calculated and shown in Fig.~\ref{fig:F2}a.~We again emphasize that the dispersion curves are immune to the coiling because of the rotational locking~\cite{willey2022coiled}.~The longitudinal response at a given junction in the coiled PSub is denoted $\eta_{i}$.~When analyzing the coiled PSub as a standalone structure, uncoupled to the flow, a complex-valued frequency response function is obtained (see Appendix).~This quantity contains information about the displacement amplitude $|\eta_{i}|$ and phase $\phi_{i}$ of the structure’s response at each junction $i$.~In the case of an uncoiled structure (0 coiling cycles), the harmonic excitation is applied at the first junctions ($i=1$) and the complex response is likewise measured at the first junction. For all subsequent coiling configurations (e.g., 1, 2, and 3 coiling cycles), the harmonic excitation is distributed uniformly among the junctions that will be interacting with the flow upon installation, thus its magnitude is divided by the number of junctions excited. For instance, in the case of three coiling cycles, the junctions 1, 3, 5, and 7 are each excited so the magnitude of the harmonic force is $F/4$ for each of these junctions.~To obtain the complex frequency response of the coiled cases, the transfer functions for each combination of excited and responsive junctions are calculated and superimposed. For instance, in the case of only 1 coiling cycle, we excite the structure at Junctions 1 and 3, and register the response also at Junctions 1 and 3; so the set of transfer functions here is (1,1), (1,3), (3,1), (3,3), where the first entry of (*,*) represents the junction of input (excitation) and the second entry represents the junction of output (response). In the case of 3 coiling cycles, 16 transfer functions are computed and superimposed, yielding $\eta_{\text{int}}=\Sigma_{i}\eta_{i}$.~The amplitude and phase of each coiled case are then calculated by taking the absolute value and angle, respectively, of this complex superposition of transfer functions. \\
\indent The super resonance phenomenon can be directly observed from the frequency response function (FRF) of the version of the PSub with three coiling cycles, as shown in the amplitude and phase plots of Figs.~\ref{fig:F2}b and~\ref{fig:F2}c, respectively.~This can be examined in reference to the lowest-frequency  resonant peak shown in the figure, which is at 278 Hz.~This will be considered our target resonance.~Compared to the reference uncoiled case, the 3-cycle coiling has shifted the first anti-resonant trough following this resonance from 555 Hz to 826 Hz, thus increasing the out-of-phase band from $\Delta f_{\text{OoP}}= 277$ Hz to $\Delta f_{\text{OoP}}= 548$ Hz, i.e., almost doubling it.~Thus, the coiling and multiple connectivity operation, as outlined above, has transformed this target resonance into a super resonance~\footnote{It should be noted that without the described multiple structural connectivity to the flow, the coiled PSub will yield an identical FRF as its uncoiled counterpart. The sole purpose of the coiling is primarily to enable the geometric convergence of the multiple junctions to a specific point in space, namely the point that interfaces with the flow. The broadening of the out-of-phase region in the FRF is enabled by the superposition of the transfer functions.}.~Upon further examination, we observe that the new 826-Hz anti-resonance trough coincides with the second resonance peak appearing in the original uncoiled structure's FRF, at practically the same frequency.~Thus, the out-of-phase region associated with the 278-Hz target resonance has effectively extended to the next anti-resonance of the coiled PSub, which is at 1737 Hz—yielding an out-of-phase band of $\Delta f_{\text{OoP}}= 1459$ Hz.~We refer to this further widened band characteristic as a quasi-super resonance, since it practically exhibits the super-resonance behavior yet stretches its out-of-phase bandwidth further.~Compared to the original configuration, the quasi super-resonant out-of-phase band is more than five times larger.\\
\indent~Figure~\ref{fig:F2}d shows the performance metric $P$ for the uncoiled and coiled structures.~This quantity is calculated as the product of the displacement amplitude and phase~\cite{Hussein_2015}.~Regions where the $P$ function is negative (green) indicate TS wave frequencies that would be stabilized by the PSub once it is employed for flow control, while regions where the $P$ function is positive (red) indicate TS wave frequencies that would experience destabilization.~As observed in Fig.~\ref{fig:F2}, as the number of coils increases in the PSub, the regions of stabilization are widened while, in contrast, the destabilization regions are narrowed$-$which is an exceptionally favorable outcome for broadband flow stabilization.~Furthermore, we observe that the negative $P$ function broadening occurs in both the width (frequency range) and intensity (magnitude), and, when the PSub is coiled three times we have entirely avoided two $P > 0$ windows within the frequency range of interest (from 250 to 1500~Hz) that would have otherwise caused flow destabilization.~Thus, effectively, the new coiled PSub design has with three cycles seamlessly broadened the bandwith of flow stabilization by a factor greater than 5, as dictated by the quasi-super resonance response .\\
\begin{figure*} [t!]
\centering
\includegraphics[width=0.7\textwidth]{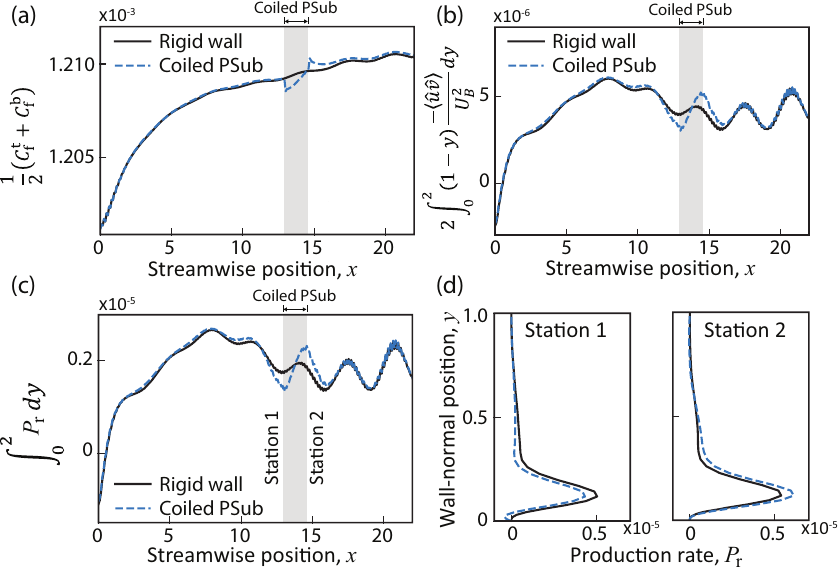}
\caption{\textbf{Super-resonant coiled PSub behavior by FIK integral analysis:}~Evaluation of various key flow parameters when the super-resonant 3-cycle coiled PSub is installed versus the rigid-wall case.~(a) Streamwise distribution of the averaged skin-friction coefficient $C_f$,
(b) streamwise evolution of the ``turbulent enhancement'' to $C_f$ computed by the FIK equation,
(c) streamwise variation of the wall-normal integral of the production rate $P_{\text{r}}$, and (d) wall-normal variation of $P_{\text{r}}$ at streamwise positions corresponding to the leading (Station 1) and trailing (Station 2) edges of the control surface.}
\label{fig:F4}
\end{figure*}

\noindent \textbf{Broadband flow stabilization} \\
\indent The broadband 3-cycle PSub configuration of Fig.~\ref{fig:F2} is installed into our channel-flow model and examined by coupled fluid-structure simulations.~Figure~\ref{fig:F3}a displays the stability map for the channel obtained by Orr-Sommerfeld stability analysis~\cite{Orr1907,Sommerfeld1909}.~The plot shows that at $Re=7500$, the unstable regime spans the frequency range 573–762 Hz.~For our selected $Re$ number we identify four frequencies corresponding to the four TS modes we consider for the subsequent coupled fluid-PSub simulations.~As listed earlier, these frequencies are 600 Hz, 650 Hz, 700 Hz, and 750 Hz, respectively.~These values are specifically selected to span the entire unstable domain.~In Fig.~\ref{fig:F3}b, we zoom into the portion of the performance metric plot that we target and identify the same four frequencies.~In Figs.~\ref{fig:F3}c and~\ref{fig:F3}d, we analyze the impact of the coiled PSub on the velocity perturbation field in the flow as obtained by DNS.~The velocity vector field is decomposed into ${\mathbf{u}} = {\langle{\bf u}\rangle} + {\mathbf {\hat{u}}}$, where ${\langle{\bf u}\rangle}$ is the mean flow component obtained by averaging ${\mathbf{u}}$ over a time range and ${\bf \hat{u}}$ is the perturbation (instability) component.~Figure~3c shows the streamwise-dependent perturbation kinetic energy, averaged over time and over the wall-normal direction,~\( K_p(x) = \int_0^{2\pi} \int_0^2 \frac{1}{2} \left( \left\langle \hat{u} \hat{u} \right\rangle + \left\langle \hat{v} \hat{v} \right\rangle + \left\langle \hat{w} \hat{w} \right\rangle \right) \, \mathrm{d}y \, \mathrm{d}z \).~We show this quantity for a case where a 3-cycle coiled PSub is installed beneath the channel covering a streamwise range of $8\le x \le 9.6$.~In this case of 3 coiling cycles, the PSub has 4 points of contact with the flow, thus the pressure-induced force exerted by the flow on each contact junction is $F/4$.~The combined (superposed) PSub's temporal response is then obtained by direct superposition, similar to the offline frequency-domain PSub characterization analysis described earlier.~Dashed lines represent $K_{\text{p}}$ in the absence of a PSub, while solid lines represent $K_{\text{p}}$ when the flow is interfaced with the coiled PSub.~We use this same PSub for all the following cases.~The four selected  distinct TS waves are simulated individually, revealing flow stabilization by the coiled PSub in all cases, with stabilization levels consistent with the performance metric predictions of Fig.~\ref{fig:F3}b.~Recall, for each TS wave, the streamwise growth rate is given by the imaginary part of the streamwise wavenumber and the strength of stabilization by the magnitude of $P$ for the 3-cycle coiled PSub at the corresponding frequency.~The results of DNS are consistent with these predictions.~We then run a simulation where all four TS waves are introduced to the flow simultaneously, with the results demonstrating stabilization of the entire set within the region of application of the coiled PSub.~It is shown that for this multi-frequency case, the total rigid-wall growth rate of $K_{\text{p}}$ is much greater, as expected, than of each individual instability; and given that four instabilities are stabilized, the level of the total local drop in $K_{\text{p}}$ exhibited by the coiled PSub is greater than any of the individual cases.~Finally, Fig.~\ref{fig:F3}d contrasts the performance of the 3-cycle coiled PSub versus the nominal uncoiled configuration for the 700-Hz instability case$-$confirming the transformation from the red destabilization zone to the green stabilization zone shown in the performance metric in (b).~In summary, the proposed coiled PSub is confirmed to be capable of stabilizing TS modes covering the entire spatial instability region of Fig.~\ref{fig:F3}b.\\

\noindent \textbf{Manipulation of turbulence production mechanism} \\
\indent A practical analysis tool to evaluate the performance of flow control schemes is through an integral equation~\cite{FIK_2,kianfar2025moment}.~To quantify the impact of the 3-cycle coiled PSub-based flow control on the flow field, we employ the Fukagata–Iwamoto–Kasagi (FIK) identity$-$an integral analysis tool derived from the second moment of the streamwise momentum equation for internal flows. This identity has been widely used to study complex internal flows and assess the effectiveness of flow control strategies.~\cite{FIK_1,FIK_2}\\
\indent~The extended FIK identity is analytically derived in this paper for spatially developing channel flows with non-zero wall boundary conditions.~The skin-friction coefficent $C_\text{f}$ in terms of this identity is expressed as
\begin{equation}
\begin{aligned}
&\frac{C_\text{f}^\text{t}+C_\text{f}^\text{b}}{2} =  {\left\{\frac{12}{Re_\text{B}}\right\}}
+ {\left\{2 \int_0^{2}\left(1 - y\right) \frac{-\langle \hat{u} \hat{v} \rangle}{U_\text{B}^2} dy\right\}}
+\\
& {\left\{ \frac{U_\text{t} V_\text{t} - U_\text{b} V_\text{b}}{U_\text{B}^2} - \frac{\langle \hat{u} \hat{v} \rangle_\text{t} - \langle \hat{u} \hat{v} \rangle_\text{b}}{U_\text{B}^2} - \frac{4}{Re_\text{B}} \frac{U_\text{t} + U_\text{b}}{U_\text{B}} \right\}}
+ {\left\{ \mathcal{I}_x \right\}},
\end{aligned}
\label{eq:FIKMain}
\end{equation}
where $Re_\text{B} = 2U_\text{B}Re$ is the Reynolds number based on bulk velocity $U_\text{B} = \frac{1}{2}\int_0^2\langle u \rangle dy$, and the superscripts “$\text{t}$” and “$\text{b}$” refer to the top and bottom walls, respectively.~The first and second terms on the right-hand side of Eq.~\eqref{eq:FIKMain} quantify the contributions of laminar and turbulent effects to the skin-friction coefficient, respectively.~Similarly, the third term quantities the effects of wall boundary conditions due to the presence of the coiled PSub.~Finally, the term $\mathcal{I}_x$ denotes the dimensionless contribution of streamwise-developing flow features to the skin-friction coefficient.~Equation~\eqref{eq:FIKMain} is written in a normalized form using $U_c$ and $\delta$.~The derivation and detailed discussion of this identity, and its application to our coiled PSub flow control problem, is given in the Appendix.\\
\indent~In the controlled cases with the PSub, the wall-shear stress is asymmetric between the top and bottom walls, which justifies using the averaged skin-friction coefficient on the left-hand side of Eq.~\eqref{eq:FIKMain}.~Figure~\ref{fig:F4}a presents this averaged skin-friction coefficient, $C_\text{f}$, as a function of the streamwise position for the 4-mode simulation scenario showing results for both the controlled (super-resonant 3-cycle coiled PSub) and reference (rigid wall) cases.~While about $98\%$ of $C_\text{f}$ is generated by the viscous laminar effects, Fig.~\ref{fig:F4}a exhibits a localized reduction in $C_\text{f}$ within the control region, confirming that the PSub interacts with the near-wall flow structures and reduces the wall-shear stress.\\
\indent~This interaction manifests through manipulation of the Reynolds shear stress ($\langle \hat{u} \hat{v} \rangle$ velocity correlation) as shown in Fig.~\ref{fig:F4}b. This plot demonstrates the quantified ``perturbation enhancement'' to friction based on the second term in Eq.~\eqref{eq:FIKMain}. At the leading edge of the control region, the coiled PSub reduces this enhancement by approximately 15\%. This weakening is strongest near the wall, where the linear weighting function (distance from the wall) in the FIK identity amplifies the contribution of near-wall perturbations. Midway through the control surface, the Reynolds shear stress recovers and eventually surpasses the reference rigid-wall case at the trailing edge, suggesting a re-energization of flow instabilities downstream of the control.\\
\indent~Figure~\ref{fig:F4}c further supports this behavior by showing the streamwise variation of the wall-normal integral of the perturbations production rate, $\int_0^2P_r dy$.
The production rate $P_{\text{r}}$, defined as $P_{\rm r}(x,y) = - \langle \hat{u} \hat{v}\rangle \frac{\partial \langle u \rangle}{\partial y}$, represents the energy transfer from the laminar base flow to the instabilities.~Since the base flow remains largely unaltered by the coiled PSub (i.e., $\partial \langle u \rangle / \partial y$ is constant), changes in $P_{\text{r}}$ are attributed directly to manipulations in $\langle \hat{u} \hat{v} \rangle$. 
Figure~\ref{fig:F4}d further explores how the coiled PSub manipulates $P_{\text{r}}$; it demonstrates the wall-normal variation of the production rate at two streamwise positions: the leading edge of the control surface (Station 1), and the trailing edge of the control surface (Station 2).~At Station 1, the observed reduction in the production rate within the control region.~In contrast, the reduction of $P_{\text{r}}$ is reversed at Station 2, explaining why $K_\text{p}$ (shown in Fig.~\ref{fig:F3}) recovers to the reference rigid wall.~These results align with prior findings.~Such production rate behavior has been observed in flow stabilization by nominal uncoiled PSub configurations~\cite{kianfar2023phononicNJP,kianfar2023local}.\\
\begin{flushleft}
\textbf{CONCLUSION}
\end{flushleft}
\indent In conclusion, this work presents the concept of \mbox{super resonance}—a new fundamental regime in resonance physics characterized by the persistence of out-of-phase modal response across a broad spectral range, despite the presence of adjacent resonant modes in the nominal configuration.~In this regime, the narrowband limitation of a target mode is effectively lifted, and the associated phase-flipping behavior with nearby modes is prevented.~This unique behavior is realized by a mechanism of multiple structural energy pathways spatially converging to a single narrow spatial region where the response is measured.~This conceptual advance opens new directions in wave-matter interaction and phase-based control strategies across physics, materials science, and engineering.

~Within this broader framework, we demonstrate a super-resonant PSub designed to significantly widen the frequency range over which TS instabilities are attenuated within a laminar channel flow.~This entails the generation of responsive vibratory motion, at an effectively single location interfacing with the flow, whereby the phase is contiguously negative over a frequency range that exceeds the characteristic bandwidth of a conventional structural resonance.~The PSub is formed from a coiled configuration with periodic rotational locking.~By enabling multiple interaction points along the coiled structure to influence that single location at the flow interface, we achieve passive stabilization across a frequency range more than five times broader than that is attainable with a corresponding uncoiled PSub.~Across this entire range, the rate of production of the perturbation kinetic energy in the flow is suppressed.~This broadened control window is shown to be sufficiently wide to fully cover, for a given Reynolds number, the unstable frequency band predicted from linear stability theory.~This approach thus greatly enhances the utility of PSubs by enabling broadband flow control—a key trait lacking across existing passive and active flow-instability control technologies~\cite{Gad2003}.~As a bonus, the total height of the structure beneath the surface is reduced, which is critical for practical implementation.~Upon integration with PSub concepts for downstream control~\cite{willey2023multi,hussein2025scatterless,klauss2025control}, super-resonant coiled PSubs will provide a robust and tunable mechanism for delaying transition to turbulence in real-world channel and boundary layer flows.~Most significantly, the concept offers a pathway towards controlling fully developed turbulent flows, whose broadband spectrum has long resisted effective control.~Reaching this target will be accelerated by further enriching the design of the constituent components of super-resonant PSubs, drawing on the extensive body of knowledge in phononic crystals and elastic metamaterials~\cite{deymier2013acoustic,hussein2014dynamics,Phani_2017,Jin_2021}, including the emerging themes of topological~\cite{ma2019topological}, nonlocal~\cite{chen2025nonlocal}, and nonlinear~\cite{fronk2023elastic} phononics, to enable additional broadening of the out-of-phase frequency range.\\
\indent The present finding marks a crucial step in the flow-phonon interaction paradigm~\cite{Hussein_2015} which targets incorporation of phononic principles into industrial flow control strategies—with far-reaching implications for air, sea, and land vehicle performance, among a wide range of other applications~\cite{avallone2025metamaterials}.\\

\noindent \textbf{Acknowledgments} \\
\indent This research is funded by Office of Naval Research Multidisciplinary University Research Initiative (MURI) Grant Number N0001421268. This work utilized the Alpine high performance computing resource at the University of Colorado Boulder. Alpine is jointly funded by the University of Colorado Boulder, the University of Colorado Anschutz, and Colorado State University and with support from NSF grants OAC-2201538 and OAC-2322260.\\

\noindent \textbf{Author contributions} \\
\indent ARH performed parametric design of coiled phononic subsurface, stability analysis, and fluid-structure simulations; AK performed turbulence production analysis; DR performed homogenization analysis; DY performed machine-learning work; CB contributed to turbulence production analysis; MIH conceived of super-resonance concept and its realization by spatially convergent multiple energy pathways and rotationally locked coiling. MIH supervised the research and wrote the paper, and ARH, AK, DR, DY, and CB edited respective sections.  

\appendix*
\section*{APPENDIX}

\renewcommand{\thesubsection}{A\arabic{subsection}}
\renewcommand{\thefigure}{A\arabic{figure}}
\renewcommand{\thetable}{A\arabic{table}}
\renewcommand{\theequation}{A\arabic{equation}}

\setcounter{figure}{0}    
\setcounter{table}{0}    
\setcounter{equation}{0}    

In this appendix, additional modeling information and analysis of the results are provided to support the main article. In Section~\ref{app:1}, we provide supplementary modeling information pertaining to the coupled fluid-structure simulations. In Section~\ref{app:2}, we overview the phononic subsurface (PSub) coiling protocol and associated beam model kinematics. Section~\ref{app:3} details the homogenization scheme and machine learning calculations referred to in the main article. These are done so the coiled PSub complex three-dimensional (3D) model is represented by a simpler one-dimensional (1D) model to enable the execution of extensive coupled flow-PSub simulations. Section~\ref{app:4} overviews key flow and PSub quantities obtained by post-processing the simulation data. Finally, in Section~\ref{app:5}, we detail the derivation of the flow integral analysis framework that allows us to decompose the various components of the skin-friction coefficient for a channel flow controlled by a coiled PSub.

\subsection{Flow-PSub modeling information}\label{app:1}
The flow simulations conducted in this investigation are direct numerical simulations (DNS) of the 3D incompressible Navier–Stokes equations in plane channel flow configurations that read
\begin{equation}
\label{eq:cont}
\frac{\partial u_i}{\partial x_i} = 0
\end{equation}
for continuity, and
\begin{equation}
\frac{\partial u_i}{\partial t} + \frac{\partial u_j u_i}{\partial x_j} = \frac{2}{Re} \delta_{ij}-\frac{\partial p}{\partial x_i} + \frac{1}{Re}\frac{\partial u_i}{\partial x_j x_j}
\label{eq:NS}
\end{equation}
for momentum conservation, respectively. The quantity $Re = U_\text{c}\delta/\nu_{\text{f}}$ is the Reynolds number based on the centerline velocity $U_\text{c}$, the half-height $\delta$ of the channel, and the fluid kinematic viscosity $\nu_{\text{f}}$.~The first term in the right-hand side of Eq.~\eqref{eq:NS}, ${2}\delta_{ij}/{Re}$, represents the mean pressure gradient driving the flow.~In the above equations, $p$ is the pressure and ${\bf u}(x,y,z,t)=(u,v,w)$ is the velocity vector with components in the streamwise $x$, wall-normal $y$, and the spanwise $z$, directions, respectively, where $t$ denotes time. Note, Eq.~\eqref{eq:NS} is in dimensionless form scaled by $U_\text{c}$ and $\delta$.

Within the coiled PSub control region, the coupling fluid-structure boundary conditions are specified as
\begin{subequations}
\label{eq:structbc}
	\begin{equation}
	\label{eq:structbcu}
	u(x_{\rm s}\le x\le x_{\rm e},y=0,z,t)=-\frac{\eta(s=0,t^*)}{\delta}\frac{du_{\rm b}}{dy}, 
	\end{equation}
	\begin{equation}
	\label{eq:structbcv}
	v(x_{\rm s}\le x\le x_{\rm e},y=0,z,t)=\frac{\dot{\eta}(s=0,t^*)}{U_\mathrm{c}}.
	\end{equation}
\end{subequations}
These are imposed to ensure the stresses and velocities match at the interface \cite{Hussein_2015}.~In Eq.~\eqref{eq:structbc}, $u_b$ denotes the base laminar streamwise velocity (i.e., the Poiseuille profile), while $\eta(s, t^*)$ and $\dot{\eta}(s, t^*)$ represent the (dimensional) displacement and velocity of the coiled PSub, respectively, with $s$ indicating the structure’s axial spatial coordinate and $t^*$ denoting the dimensional time.
Referred to as transpiration boundary conditions~\cite{Lighthill_1958,Sankar_1981}, Eqs.~\eqref{eq:structbcu} and~\eqref{eq:structbcv} are obtained by keeping the interface location fixed and retaining only the linear terms following a Taylor series expansion of the exact interface compatibility conditions. Equations~\eqref{eq:structbc} are valid if the coiled PSub motion is only in the wall-normal direction and $\eta(s=0,t^*) \ll \delta$. Hence, during DNS the roughness Reynolds number is monitored and maintained below 25\cite{morkovin1990roughness}. \\
\indent The Navier-Stokes equations are integrated using a time-splitting scheme~\cite{Dana91,saiki1993spatial,kucala2014spatial} on a staggered structured grid system.~A two-node finite-element (FE) model is used for determining the coiled PSub nodal axial displacements, velocities, and accelerations~\cite{HusseinJSV06} where time integration is implemented simultaneously with the flow simulation using an implicit Newmark algorithm \cite{Newmark}.~The FE discretization is such that each unit cell is modeled by three finite elements.~Since the equations for the fluid and the PSub are advanced separately in the coupled simulations, a conventional serial staggered scheme~\cite{Farhat_2000} is implemented to couple the two sets of time integration schemes.~This approach has been extensively verified and produced excellent agreement with the experimentally validated linear theory, giving a maximum deviation of 0.05\% in the predicted perturbation energy
growth~\cite{kucala2014spatial}.~More details on the computational models and numerical schemes used are detailed in Ref.~\cite{kianfar2023phononicNJP}. \\
\indent As described in Section C, the full-range coiling of the PSub along with the application of rotational locking at the turning segments preserves the phonon band structure.~Therefore, since we are only interested in longitudinal motion along the axial coordinate of the PSub structure, it suffices to model the coiled PSub as a rod model in our simulations.~Thus, the coiled PSub axial displacement, velocity, and acceleration are obtained by solving the governing equation for a 1D linear elastic slender rod structure with a periodic array of spring-mass resonators.~The elastodynamic equations for the backbone elastic rod are
\begin{equation}
    \rho_{\rm s} \ddot{\eta}=(E\eta_{,s}+C\dot{\eta}_{,s})_{,s}+f,
    \label{eq:structure}
\end{equation}
 where $\rho_{\rm s}=\rho_{\rm s}(s)$, $E=E(s)$, and $C=C(s)$, respectively, represent density, elastic modulus, and damping.~Equation~\eqref{eq:structure} is modified approppriately to incorprate the spring-mass resonators distributed in the manner described in the main article.~The external forcing $f$ is applied either artificially in the pre-simulation coiled PSub characterization analysis or imposed by the dimensional spanwise-spatial-average flow pressure within the control region during the coupled fluid-structure simulations.~In Eq.~\eqref{eq:structure}, the differentiation with respect to position is indicated by $(.)_{,s}$, and the superposed single dot $\dot{(.)}$ and double dot $\ddot{(.)}$ denote the first and second-time derivatives, respectively. Free-fixed boundary conditions are applied on the coiled PSub, with the forcing is applied to the free (top) end at $s=0$ where the structure interfaces with the flow.\\ 
 \indent During the coupled fluid-structure simulations, the pressure field acting on the surface over the coiled PSub is identified at each time step to yield a value for $f$.~This forcing value is applied as an input to the coiled PSub and the model is solved at each time step to produce the responding velocity as a function of $s$ within the domain of the coiled PSub.~The appropriate  velocity is then applied to the flow across the region where each junction of the coiled PSub interfaces with the flow, as described in main article.

\subsection{PSub coiling with periodic rotational locking}\label{app:2}
\begin{figure}
\centering
\includegraphics[width=0.45\textwidth]{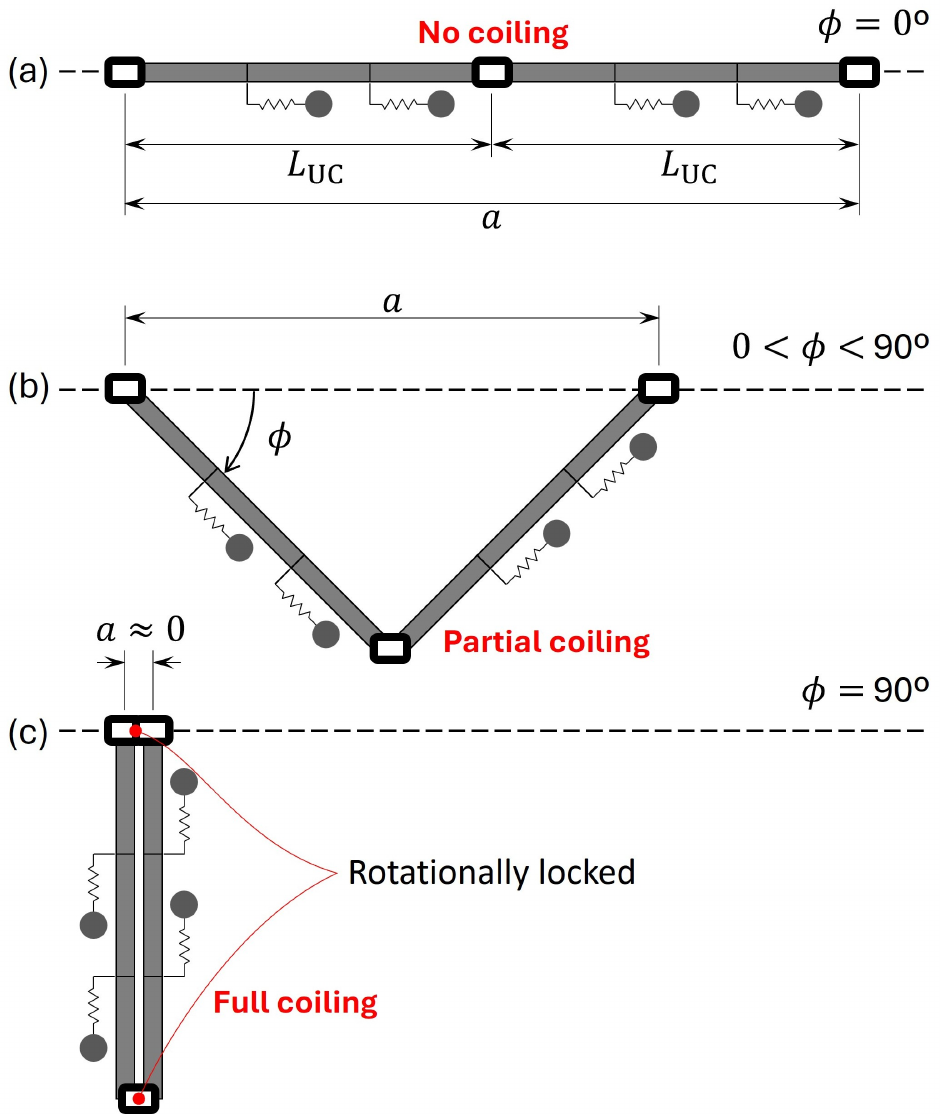}
\caption{\textbf{PSub coiling:} PSub periodic unit cell with different degrees of coiling: (a) uncoiled case, (b) partially coiled case, (c) fully coiled case. The periodic unit-cell size for the dispersion analysis scaling~\cite{willey2022coiled} is taken as $a = 2 L_\text{UC}\cos{\phi}$.}
\label{fig:B1}
\end{figure}

The PSub coiling with full 180$^\circ$ turns not only enables multiple points of interaction with the flow$-$which ultimately leads to the enhanced broadband performance$-$but it also allows us to characterize its response with the simple model of the uncoiled configuration as stated in Section A.~To demonstrate this, we can examine the dispersion diagram of a periodic substructure of the PSub composed of two unit cells at different degrees of coiling as shown inFig.~\ref{fig:B1}.~As shown, a full  180$^\circ$ turn corresponds to an angle of $\phi=90^\circ$ when considering the rotation angle of the coiling elements relative to the initial configuration.~In order to take into account the two-dimensional response of the partially coiled cases, a standard finite element beam model is considered, in which a nodal rotation degree of freedom $\theta$, associated with bending deformations, is admitted in addition to the nodal displacements in the longitudinal and transversal directions $\eta$ and $\zeta$, respectively.~The bending response in the transversal direction is driven by
\begin{equation}
    \rho_\text{s}\ddot{\zeta}+ (EI \theta_{,s})_{,s} = 0,
    \label{eq:bending}
\end{equation}
where $\theta=\zeta_{,s}$, and $I$ is the area moment of inertia normalized by a reference cross-sectional area.
~To make the analysis consistent with the PSub design proposed in the main text, the same properties have been considered here.~With the cross-sectional side length defined as $w_{\text{c}}$, the area moment of inertia is defined as $I = w_{\text{c}}^2/12$, under the assumption that Eq.~\eqref{eq:bending} has been normalized by the same reference area as Eq.~\eqref{eq:structure}. 

\begin{figure*}[t!]
\centering
\includegraphics[width=0.9\textwidth]{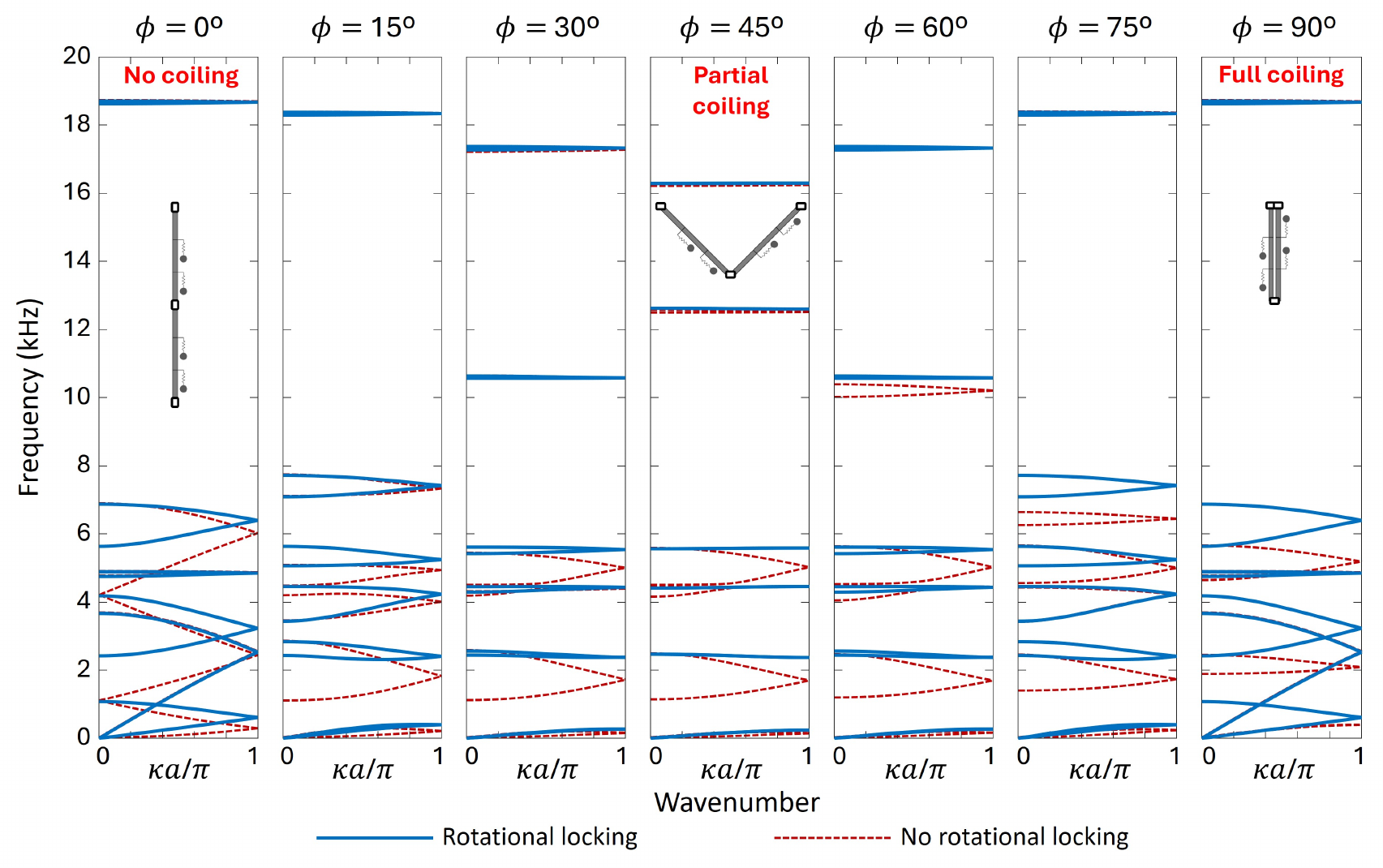}
\caption{\textbf{Conservation of phonon dispersion with coiling:} Phonon and structure of PSubs for different degrees of coiling $\phi$ resulting from applying rotational locking at the coiling junctions (solid blue curves) versus leaving the rotation degree of freedom free (dashed orange curves).~At the extreme cases, for $\phi = 0^\circ$ (uncoiled) and $\phi = 90^\circ$ (fully coiled), the phonon band structure is preserved only when rotational locking is applied.}
\label{fig:B2}
\end{figure*}

In Ref.~\cite{willey2022coiled}, it was demonstrated that the phonon band structure in the uncoiled and fully coiled cases is preserved when the rotation degree of freedom is locked at the coiling junctions, i.e., $\theta = 0$ at the associated nodes.~The same principle applies to the PSub design described in the present work, as it can be observed in Fig.~\ref{fig:B2}, which shows the full phonon band structures obtained for different degrees of coiling, with and without rotational locking.~By comparing the uncoiled ($\phi=0^\circ$) and fully coiled ($\phi=90^\circ$) cases, it can be seen that when rotational locking is applied: (1) the resulting dispersion diagrams are exactly identical and (2) the longitudinal and transversal degrees of freedom are uncoupled. Details on how the rotational locking can be realized experimentally in practice can be found in Ref.~\cite{willey2022coiled}.

As a consequence of this result, as long as rotational locking is applied at the coiling junctions, one can assess the PSub's performance metric of the fully coiled case through a model of the uncoiled configuration. Furthermore, since the frequency response function is obtained for longitudinal excitations of the PSub structure (in both the uncoiled and fully coiled cases), Eq.~\eqref{eq:bending} is not necessary and one can simply consider Eq.~\eqref{eq:structure}.

\begin{figure*}[t!]
\centering
\includegraphics[width=0.9\textwidth]{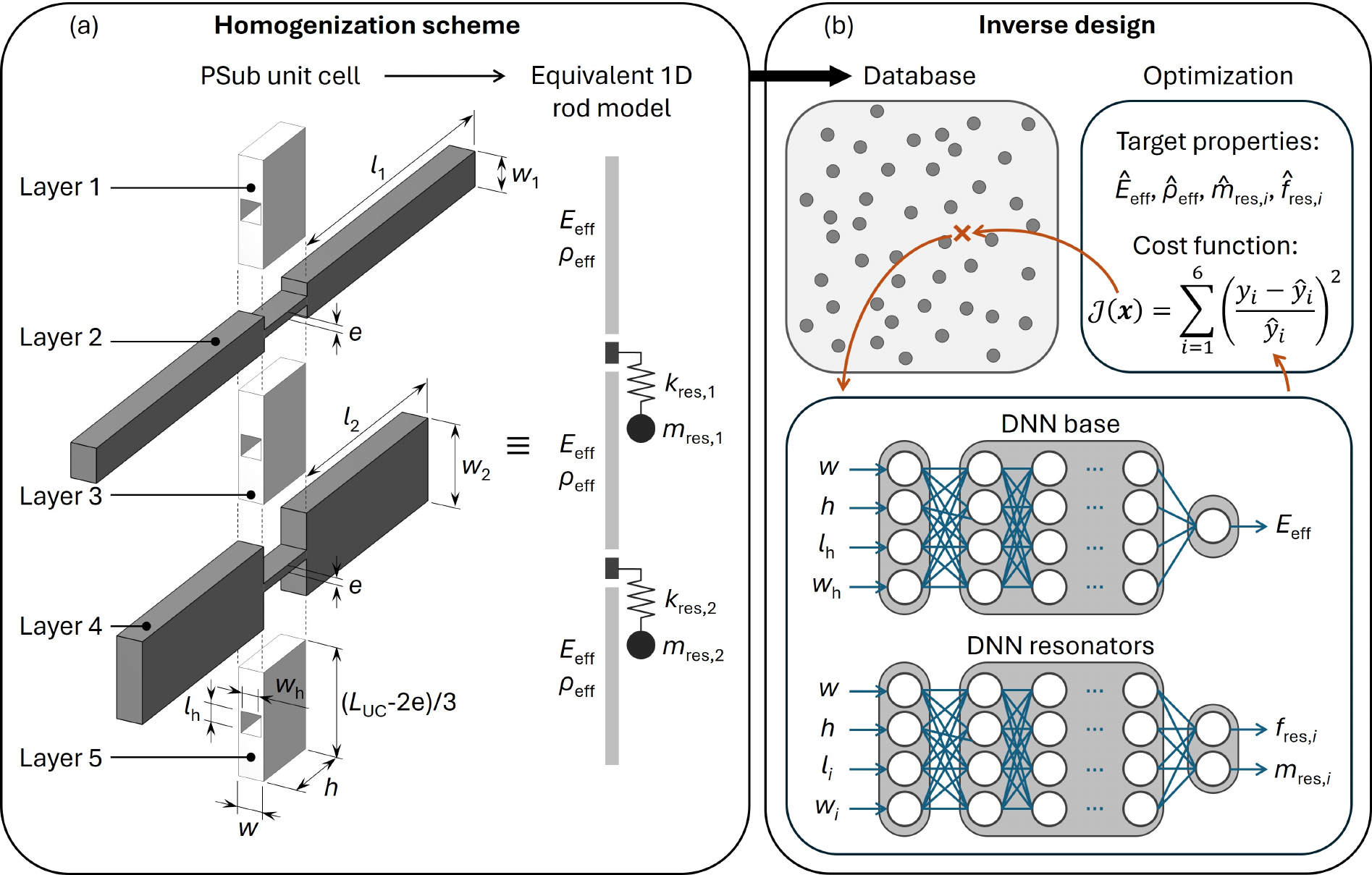}
\caption{\textbf{Schematic representation of the homogenization and machine-learning based inverse design approach:} (a) The coiled PSub unit cell design (on the left) is split into different layers, distinguishing between the base component (Layers 1, 3 and 5) and the resonators (Layers 2 and 4). Each layer is parametrized with relevant geometrical features. For a given set of parameters, the response of the coiled PSub unit cell design can be studied with the equivalent 1D rod model on the right, in terms of the depicted effective properties obtained from the homogenization scheme. (b) For a target set of properties that provide the intended broadband performance, the inverse design approach allows us to obtain a set of geometrical parameters of the coiled PSub unit cell that trigger the equivalent response.~The process involves building databases for each type of layer with randomized combinations of parameters, then the use of deep neural networks to create surrogate models that enable interpolation within the databases.~These models are used in a GA optimization algorithm to minimize the $L_2$-norm of the relative error between the interpolated and target effective properties.}
\label{fig:homog_ml}
\end{figure*}

\subsection{Homogenization and machine-learning approach for design and reduce-order modeling of coiled PSub}\label{app:3}
The proposed coiled PSub unit-cell design depicted in Fig.~1 of the main article has been obtained following an inverse design approach that utilizes a homogenization scheme in combination with a genetic algorithm (GA) optimization process enhanced by machine learning.~Each unit cell has been split into 5 layers along the wave-propagating dimension, as depicted in Fig.~\ref{fig:homog_ml}.~Layers 1, 3 and 5 represent the base rod component, and each contains a hole extending along the spanwise dimension to reduce the effective stiffness and total mass of the structure.~Layers 2 and 4 are very thin ($e=0.5$ mm) and each one supports two  pillars extruding along the spanwise dimension.~These pillars can vibrate as cantilevered beams at specific (distinct) frequencies, acting as internal resonators that add local resonance effects to the PSub structure.~Considering that the relevant dynamic and deformation phenomena occur along the direction of wave propagation, the homogenization framework in Ref.~\cite{sal2023optimal} can be applied to this configuration to obtain an equivalent 1D model capable of reproducing the unit-cell response in terms of the displacements.~The resulting homogenized model is composed of standard rod-like elements$-$with effective elastic modulus and density properties$-$and spring-mass units attached to account for the local resonance effects (see Fig.~\ref{fig:homog_ml}).~As a result of this homogenization process, for a given 3D unit cell design, one can use Eq.~\eqref{eq:structure} to study the PSub response in terms of the effective 1D properties.~The mathematical details of the homogenization scheme can be found in Ref.~\cite{sal2023optimal}.

Once the 1D PSub model, under full coiling and periodic rotational locking is designed to produce the broad frequency response function shown in Fig.~2 of the main article, an inverse design problem is solved to obtain a corresponding 3D coiled PSub design.~This process utilizes the homogenization mapping and is conducted by means of an optimization process.~For this purpose, we utilize the methodology presented in Ref.~\cite{yago2024machine}.~First, the homogenization model is applied to each distinct unit cell configuration to build databases linking their main geometrical features to the resulting effective properties.~In the case of the base component, this is the effective modulus $E_\text{eff}$, while for the layers with resonators, these are the equivalent mass $m_{\text{res},i}$ and the resonance frequency $f_{\text{res},i}$.~The effective density of the base material $\rho_\text{eff}$ has been considered the average density of the central ``trunk'' of the whole unit cell (thus including the mass of Layers 1, 3 and 5 and also the mass of the non-resonating Layers 2 and 4).~To obtain the correct effective values, the area used for normalization is not the actual cross-sectional area of the PSub central rod component, but the area of the surface in contact with the flow, denoted $A$ in Fig.~1 of the main article.~Such area can be chosen arbitrarily, with the only restriction being that the total width in the streamwise direction, $w_\text{f}$, should be much smaller than the wavelength of the TS wave, $\lambda_\text{TS}$. In our simulations, $\lambda_\text{TS} \approx 4.1$~mm, and $w_\text{f}=0.4$~mm has been chosen. Then, the reference area is given by $A = w_\text{f}h_\text{total}$, with $h_\text{total} = h + \max{\{l_1,l_2\}}$ selected to be the total length of the PSub in the spanwise direction.

Each database is built considering 15000 sample points scattered according to a quasi-random Sobol sequence.~Then, a Deep Neural Network (DNN) is applied to obtain surrogate models of each layer enabling an interpolation of the effective properties for any given input set of parameters.~The layers with resonators are modeled using a feedforward neural network comprising five hidden layers, each containing 40 neurons and employing the swish activation function.~In contrast, the base component is represented by a deeper architecture with seven hidden layers, where the first three layers consist of 50 neurons each, while the remaining four contain 30 neurons, all utilizing the swish activation function to ensure smooth nonlinearity across the surrogate model.~All surrogate models are trained using Matlab’s built-in Adam optimizer to minimize the mean squared error (MSE) loss function, considering $L_2$ weight regularization.~Training is performed using mini-batches of 256 samples, with a piecewise constant learning rate schedule, starting at 0.01 and reduced an order of magnitude every 10 epochs.~The dataset is partitioned with 85\% of the samples allocated for training and the remaining 15\% reserved for validation to monitor performance during training.

The last step consists of employing an optimization algorithm to find the combination of geometric features that provide the closest point in the surrogate model to the specified target effective 1D parameters.~The geometric variables include the width of the structure $w$, the thickness of the base component $h$, the fractional dimensions of the internal hole in the base component $\gamma_l = 3l_\text{h}/(L_\text{UC}-2e)$ and $\gamma_w = w_\text{h}/w$, the length of the two resonant cantilever beams $l_1$ and $l_2$, and their respective thicknesses $w_1$ and $w_2$.~The values of $L_\text{UC}$ and $e$ are fixed to 20~mm and 0.5~mm, respectively.~A hybrid two-stage optimization strategy is employed, combining first a standard GA used to explore the design space globally and identify a promising candidate solution, followed by a local refinement using a constrained Sequential Quadratic Programming (SQP) algorithm to improve convergence towards the optimum.~The GA operates within specified bounds for each design variable, defined as $\mathbf{x}=\{w, h, \gamma_l, \gamma_w, l_1, l_2, w_1, w_2\}$. To account for manufacturing amenability, lower and upper bounds for each of the input parameters are set to $\mathbf{x}_\text{min}=\{1~\text{mm},1~\text{mm},0.1,0.1,5~\text{mm},5~\text{mm},0.75~\text{mm},0.75~\text{mm}\}$ and $\mathbf{x}_\text{max}=\{5~\text{mm},25~\text{mm},0.9,0.9,50~\text{mm},50~\text{mm},5~\text{mm},5~\text{mm}\}$, respectively.~The optimization objective is to minimize a cost function that quantifies the discrepancy between the surrogate model's prediction and the target effective parameters, defined as
\begin{equation}
\mathcal{J}(\mathbf{x}) = \sum_{i=1}^6\left(\frac{y_i(\mathbf{x}) - \hat{y}_i}{\hat{y}_i}\right)^2,
\end{equation}
where $\mathbf{y}(\mathbf{x}) = \{E_\text{eff}(\mathbf{x})$, $\rho_\text{eff}(\mathbf{x})$, $\mu_{\text{res},1}(\mathbf{x})$, $\mu_{\text{res},2}(\mathbf{x})$, $f_{\text{res},1}(\mathbf{x})$, $f_{\text{res},2}(\mathbf{x})\}$ are the values obtained from the DNN surrogate model for a given set of parameters $\mathbf{x}$, with $\mu_{\text{res},i} = m_{\text{res},i}/\rho_\text{eff}L_\text{UC}$. Setting the target properties to the ones providing the desired outcome, in terms of the performance metric of the equivalent 1D model, i.e., $\hat{\mathbf{y}} = \{0.2$~GPa, 450~kg/m$^3$, 3, 5, 5000~Hz, 20000~Hz$\}$, the PSub design proposed in the main article is obtained.

While the GA manages to obtain a practically exact match between $\mathbf{y}$ and $\hat{\mathbf{y}}$, it does so using the DNN surrogate models' predicted values.~This means that the actual effective properties corresponding to the optimal set of geometric parameters are not strictly the same as the target values.~For this case, the actual values would be $\mathbf{y}_\text{homog} = \{0.205$ ~GPa, 449.99~kg/m$^3$, 2.983, 4.981, 4996.68, 20524.19$\}$.~This represents a relative deviation of only 1.4\%, which is an indication of the high effectiveness of the DNN-based interpolation.

\subsection{Flow and PSub quantities obtained by post-processing the simulations data}\label{app:4}
Throughout the main article, several quantities of interest are calculated by post-processing the time-dependent data emerging from the coupled fluid-structure simulations. \\
\indent On the flow side, we decompose the velocity vector in Eqs.~\eqref{eq:cont} and \eqref{eq:NS} into ${\mathbf{u}} = {\langle{\bf u}\rangle} + {\mathbf {\hat{u}}}$, and the pressure field in Eq.~\eqref{eq:NS} into  $p = {\langle{p}\rangle} + \hat{p}$, where $\left(\langle \cdot \rangle\right)$ and $\left(\hat{\cdot}\right)$ represent time-averaged and perturbation quantities, respectively.~With this decomposition, we calculate the streamwise position-dependent integral of the perturbation kinetic energy, which is defined as
\begin{equation}
K_p\left(x\right)= \int_0^{2\pi} \int_0^{2} \frac{1}{2}\left(\left\langle\hat{u}\hat{u}\right\rangle+\left\langle\hat{v}\hat{v}\right\rangle+\left\langle\hat{w}\hat{w}\right\rangle\right) \mathrm{d} y \mathrm{~d} z,
\label{eq:Kp}
\end{equation}
Along the surface, we are interested in the wall shear stress. The wall shear stress normalized by the fluid density $\rho_f$ is defined as
\begin{equation}
\frac{\tau_w}{\rho_f} = \left[\nu_{\text{f}} \frac{\partial \langle u \rangle}{\partial y} -\langle \hat{u} \hat{v} \rangle\right]_{y=0,2}, 
\end{equation}
in which on the right-hand side the first term represents the (molecular) viscous shear stress, and the second term represents the perturbation (Reynolds) shear stress at the wall. Note, the second term is naturally zero for the rigid wall, whereas within the control region, even though substantially smaller than the viscous shear stress, it is non-zero due to the coiled PSub's motion.
We also compute the rate of production of perturbation kinetic energy~\cite{Prandtl1921,Cossu_2004}, which in dimensionless form is 
 \begin{equation}
    P_{\rm r}(x,y) =  - \langle \hat{u} \hat{v}\rangle \frac{\partial \langle u \rangle}{\partial y}.
\end{equation}

\indent On the PSub side, a relevant quantity is the elastodynamic energy within the PSub elastic domain, which is specified as
\begin{equation}
     \Psi(s, t)=\frac{1}{2}\left[E(\mathrm{~d} \eta / \mathrm{ds})^2+\rho_s \dot{\eta}^2\right].
\end{equation}
This quantity has been evaluated in prior work~\cite{kianfar2023phononicNJP}.

\subsection{Derivation of integral analysis: FIK identity for spatially developing channel flows}
\label{app:5}
This section presents the full derivation of the Fukagata–Iwamoto–Kasagi (FIK) identity~\cite{FIK_1}, with an emphasis on extending the formulation to spatially developing internal flows interacting with a coiled PSub (or any type of phononic subsurface in general), featuring non-zero wall boundary conditions (i.e., wall transpiration).

The FIK identity is derived by performing a wall-normal integration of the second moment of the governing equations, equivalent to a triple integration under Cauchy’s formula \cite{Bannier_2015}, $\int_0^{2\delta}(y-\delta)^2\left[\cdot\right]dy$. Applying this procedure to the statistically stationary, Reynolds-Averaged Navier–Stokes (RANS) streamwise ($i = 1$) momentum equation for plane channel flows gives
\begin{equation} 
\int_0^{2\delta} (y-\delta)^2
    \left[\frac{\tau_w}{\delta} - 
    \frac{\partial \langle u \rangle \langle v \rangle}{\partial y} - 
    \frac{\partial \langle \hat{u} \hat{v} \rangle}{\partial y} + 
    \nu_{\text{f}} \frac{\partial^2 \langle u \rangle}{\partial y^2} + 
    I_x\right]dy =0,
    \label{eq:RANS}
\end{equation}
where $I_x$ represents the streamwise development terms, defined as
$$
I_x = \nu_{\text{f}} \frac{\partial^2 \langle u \rangle}{\partial x^2} - \frac{\partial \langle u \rangle \langle u \rangle}{\partial x} - \frac{\partial \langle \hat{u} \hat{u} \rangle}{\partial x}.
$$
Note that the wall-normal integration is carried out from $y = 0$ to $y = 2\delta$, thereby relaxing the assumption of centerline symmetry typically used in canonical FIK derivations. 

The following provides a detailed evaluation of the integrals in the bracketed terms of Eq.~\eqref{eq:RANS}. The first term corresponds to the (constant) mean pressure gradient that drives the flow. Since the wall shear stress $\tau_w$ generated by the mean pressure gradient is a constant, this integral simplifies to
\begin{gather*}
    \int_0^{2\delta} (y - \delta)^2 \frac{\tau_w}{\delta} \, dy 
    = \frac{\tau_w}{\delta} \int_0^{2\delta} (y - \delta)^2 \, dy 
    = \left( \delta^2 U_\text{B}^2 \right) \frac{C_\text{f}}{3},
\end{gather*}

\noindent where $U_\text{B} = \frac{1}{2}\int_0^{2\delta}\langle u \rangle dy$ is the bulk velocity and $C_\text{f} = {2 \tau_w}/{\rho_f U_\text{B}^2}$ is the skin-friction coefficient.

The second term represents the contribution from mean advection. Applying integration by parts gives
\begin{gather*}
    \int_0^{2\delta} (y - \delta)^2 \frac{\partial \langle u \rangle \langle v \rangle}{\partial y} dy 
    = \left( \delta^2 U_\text{B}^2 \right) \frac{U_\text{t} V_\text{t} - U_\text{b} V_\text{b}}{U_\text{B}^2},
\end{gather*}
where the subscripts ``$\text{t}$'' and ``$\text{b}$'' refer to the values of $\langle u \rangle$ and $\langle v \rangle$ at the top and bottom walls, respectively.

The third term accounts for the influence of the Reynolds shear stress on wall-normal momentum transport. Similar integration of the second moment yields
\begin{gather*}
    \int_0^{2\delta} (y - \delta)^2 \frac{\partial \langle \hat{u} \hat{v} \rangle}{\partial y} \, dy 
    = \\
    \left( \delta^2 U_B^2 \right) \left\{ 
    \frac{\langle \hat{u} \hat{v} \rangle_\text{t} - \langle \hat{u} \hat{v} \rangle_\text{b}}{U_\text{B}^2} 
    + \frac{2}{\delta} \int_0^{2\delta} \left(1 - \frac{y}{\delta} \right) \frac{-\langle \hat{u} \hat{v} \rangle}{U_\text{B}^2} \, dy 
    \right\}.
\end{gather*}
It is important to note that the Reynolds shear stress at the walls, $\langle \hat{u} \hat{v} \rangle_\text{t}$ and $\langle \hat{u} \hat{v} \rangle_\text{b}$, are not necessarily zero, depending on the wall boundary conditions. For example, low-amplitude vibration of the control surface in the coiled PSub configuration induces non-zero $\langle \hat{u} \hat{v} \rangle$ at the bottom wall. 

The fourth term in Eq.~\eqref{eq:RANS} represents viscous diffusion. Evaluating its integral using integration by parts twice yields
\begin{gather*}
    \int_0^{2\delta} (y - \delta)^2 \left(\nu_{\text{f}} \frac{\partial^2 \langle u \rangle}{\partial y^2} \right) dy 
    = \\
    (\delta^2 U_\text{B}^2) \left\{ 
    \frac{8}{Re_\text{B}}-C_\text{f} - \frac{4}{Re_\text{B}} \left( \frac{U_\text{t} + U_\text{b}}{U_\text{B}} \right) 
    \right\},
\end{gather*}
where $Re_\text{B} = {2\delta U_\text{B}}/{\nu_{\text{f}}}$ is the Reynolds number based on the bulk velocity~\cite{Pope}.
This expression highlights the explicit contributions of the wall shear stress (through $C_\text{f}$) and the bulk flow scaling (through $Re_\text{B}$), thereby offering a physical rationale for taking the second moment of the streamwise momentum equation in internal flows, as done in the FIK identity~\cite{FIK_1}.

The final term on the left-hand side of Eq.~\eqref{eq:RANS} accounts for contributions from spatial development of the flow. 
Taking the wall-normal integral of the second moment of this term yields
\begin{gather*}
    \int_0^{2 \delta}(y-\delta)^2 I_x \, dy = U_\text{B}^2 \delta^2 \int_0^{2 \delta}\left(1 - \frac{y}{\delta}\right)^2 \frac{I_x}{U_\text{B}^2} \, dy =\\ (\delta^2 U_\text{B}^2) \mathcal{I}_x.
\end{gather*}
Here, $\mathcal{I}_x$ denotes the dimensionless contribution of streamwise-developing flow features. 
While $\mathcal{I}_x$ vanishes in the limit of fully developed internal flows, it is non-zero in spatially evolving configurations. 
For example, in this work, the flow exhibits growth of instability waves [e.g., Tollmien–Schlichting (TS) modes] along the streamwise direction. 
Moreover, passive actuation via the coiled PSub control mechanism further amplifies spatial variation in the flow field. 
Accordingly, the extended FIK identity presented here explicitly retains this contribution to ensure a comprehensive and robust integral analysis.

By gathering and re-arranging the integrals of all the terms described above and normalizing by the square of the outer length and velocity scales, $(\delta U_\text{B})^2$, the FIK identity takes the form
\begin{equation}
\begin{aligned}
&C_\text{f} =  \underbrace{\left\{\frac{12}{Re_{\text{B}}}\right\}}_{I_{\text{Lam}}}
+ \underbrace{\left\{\frac{2}{\delta} \int_0^{2 \delta}\left(1 - \frac{y}{\delta}\right) \frac{-\langle \hat{u}  \hat{v} \rangle}{U_\text{B}^2} dy\right\}}_{I_{\text{Turb}}}
+
\underbrace{\left\{ \mathcal{I}_x \right\}}_{I_{\text{Stream}}}\\
& + \underbrace{\left\{ \frac{U_\text{t} V_\text{t} - U_\text{b} V_\text{b}}{U_\text{B}^2} - \frac{\langle  \hat{u}  \hat{v} \rangle_\text{t} - \langle  \hat{u}  \hat{v} \rangle_\text{b}}{U_\text{B}^2} - \frac{4}{Re_\text{B}} \frac{U_\text{t} + U_\text{b}}{U_\text{B}} \right\}}_{I_{\text{PSub}}},
\end{aligned}
\label{eq:FIK}
\end{equation}
where each term in braces $\{\cdot\}$ corresponds to a distinct flow feature contributing to the skin-friction coefficient $C_\text{f}$ on the left-hand side. The left-hand side form of Eq.~\eqref{eq:FIK} assumes equal skin friction at the bottom and top walls. This assumption can be further relaxed by substituting the left-hand side with
$\frac{1}{2}\left(C_\text{f}^\text{t} + C_\text{f}^\text{b}\right)$, as presented in the main text by Eq. 1. This decomposition provides a quantitative mapping of how various flow features impact the overall skin-friction coefficient. 

The following provides a physical interpretation of each term in the FIK identity with respect to their assigned number in Eq.~\eqref{eq:FIK}: 
\begin{itemize}
    \item $I_{\text{Lam}}$--\textit{Laminar friction:} This term is inversely proportional to the Reynolds number and quantifies what the skin-friction coefficient would be if the flow was laminar at the same Reynolds number. While the actual flow might be transitional or turbulent, this provides a baseline for comparison.
    \item $I_{\text{Turb}}$--\textit{Turbulent (or perturbation) friction} This term quantifies the explicit contribution of Reynolds shear stress to friction drag. Its linear weighting by wall-normal distance reflects the use of the second moment in the integral formulation.
    \item $I_{\text{Stream}}$--\textit{Streamwise-development friction:} This term accounts the flow effects due to spatial development in the streamwise direction, i.e., non-homogeneity in the streamwise direction.
    \item $I_{\text{PSub}}$--\textit{Wall boundary-condition friction:} This term captures the influence of imposed non-zero boundary conditions—both mean and fluctuating—at the walls. For example, in flows controlled by coiled PSub vibrations, surface-imposed velocity fluctuations contribute to this term. 
\end{itemize}

\begin{figure} [t!]
\centering
\includegraphics{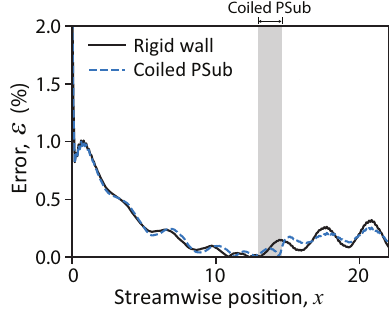}
\caption{\textbf{FIK error analysis:} Relative error in the extended FIK identity along the streamwise direction for the multi-mode TS wave case. The dashed lines denote the bounds of the super-resonant 3-cycle coiled PSub control surface.}
\label{fig:D1_error}
\end{figure}
\begin{figure*}[t!]
\centering
\includegraphics[width=0.7\textwidth]{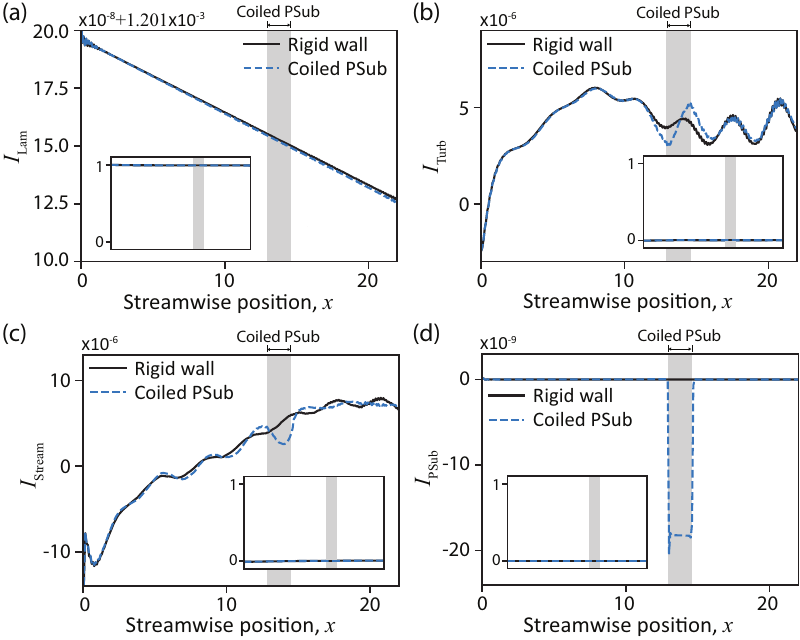}
\caption{\textbf{FIK dissection of skin-friction coefficient:} Contributions to the skin-friction coefficient of individual flow features from the right-hand side of the extended FIK identity given in Eq.~\eqref{eq:FIK} as a function of streamwise position; a) laminar friction, b) perturbation (or turbulent) friction, c) streamwise-development friction, and d) wall boundary-condition friction. Insets show the normalized contributions with respect to the total skin-friction coefficient.}
\label{fig:D2_FIK}
\end{figure*}

\subsection{Budget of the extended FIK identity}\label{app:6}
This section quantifies the contribution to the skin-friction coefficient of all the flow features characterized by the extended FIK identity given by Eq.~\eqref{eq:FIK}. 

Figure~\ref{fig:D1_error} shows the relative error associated with the FIK identity along the streamwise direction for the case involving multiple TS waves controlled by the 3-cycle coiled PSub (see main article).~This relative error is defined as
\begin{equation}
\epsilon = \frac{|\text{LHS} - \text{RHS}|}{\text{LHS}} \times 100,
\end{equation}
where LHS and RHS refer to the sum of the left- and right-hand sides of Eq.~\eqref{eq:FIK}, respectively. The error remains below 2\% throughout the domain, with an average of less than 0.5\%. This small discrepancy confirms the accuracy and robustness of the extended FIK identity, particularly in quantifying the influence of coiled PSub vibrations on the flow field and hence the skin-friction coefficient.

Given the demonstrated accuracy, Fig.~\ref{fig:D2_FIK} presents the complete budget of the extended FIK identity for both the uncontrolled (rigid wall) and controlled (super-resonant 3-cycle coiled PSub) cases. This figure explicitly illustrates the contributions of individual flow features in the extended FIK identity, Eq.~\eqref{eq:FIK}, to the left-hand side, i.e., the skin-friction coefficient.

The first observation reveals that the order of magnitude of these contributions is extremely disparate. For instance, the base ``laminar friction'' is on the order of $10^{-3}$ and approximately equal to the average of the top and bottom wall skin-friction coefficients (left-hand side of Eq.~\eqref{eq:FIK}). In contrast, the explicit contribution from flow instabilities  (i.e., the perturbation enhancement) is on the order of $10^{-6}$. This disparity is consistent with the flow regime examined. Although subjected to unstable TS waves, the flow is still in the early stages of transition and behaves predominantly as a laminar channel flow.

Figure~\ref{fig:D2_FIK}a shows the variation of the laminar friction term along the streamwise direction, with the inset presenting its normalized contribution relative to $C_\text{f}$. This term is inversely proportional to the bulk velocity, consistent with analytical relations~\cite{Pope}. The plot indicates a weak but monotonic increase in $U_\text{B}$ downstream as instabilities amplify. The presence of the coiled PSub control does not alter this trend. As shown in the inset, the laminar contribution accounts for nearly total friction drag.

The direct contribution of instabilities is depicted in Fig.~\ref{fig:D2_FIK}b, identical to the plot showne in Fig.~4b in the main article.~The perturbation-induced wall-normal momentum transport, and hence its contribution to friction drag (i.e., perturbation enhancement), is minimal. 
Moreover, whereas the coiled PSub did not impact the laminar friction, it visibly affects the flow perturbations, particularly $\langle  \hat{u} \hat{v} \rangle$, within the control region. 
The strength of the coiled PSub response is anticipated to correlate with the strength of the flow perturbations: stronger fluctuations would elicit more significant control effects. 

Figure~\ref{fig:D2_FIK}c presents the contribution from the streamwise-developing term, which comprises several sub-terms. Among these, $\partial \langle \hat{u} \hat{u} \rangle/\partial x$, representing the streamwise velocity RMS gradient, is the dominant component. Although the streamwise inhomogeneity terms are small (on the order of $10^{-6}$ due to the imposed fully developed Poiseuille base flow), their contribution is comparable to that of the ``turbulent enhancement'' term. Similar to its effect on Reynolds shear stress, the coiled PSub also interacts with $\langle \hat{u} \hat{u} \rangle$, resulting in an approximately 80\% reduction in $\mathcal{I}_x$ within the control region. Downstream of the control surface, $\mathcal{I}_x$ recovers to values observed in the rigid wall case.

As discussed in Section E, the extended FIK identity includes an explicit term that captures the effect of non-zero wall boundary conditions on the skin-friction coefficient. In this study, the coiled PSub imposes such conditions along the control surface region; this contribution is denoted by ${I}_{\text{PSub}}$ and shown in Fig.~\ref{fig:D2_FIK}d.~Following the fluid-structure simulation conditions from Ref.~\cite{Hussein_2015}, the roughness Reynolds number at the control surface is monitored to ensure a hydraulically smooth wall. That is, the coiled PSub's vibrations in response to flow forcing remains infinitesimal.~Consequently, ${I}_{\text{PSub}}$ contributes negatively to $C_\text{f}$ (indicating stabilization), but its magnitude is negligible relative to other terms. Larger coiled PSub displacements would likely lead to more significant contributions from this term.

Overall, this section provides a complete budget analysis of the extended FIK identity derived in this work (Eq.~\eqref{eq:FIK}), quantifying the absolute and relative contributions of distinct flow features to the skin-friction coefficient. The minimal error observed in the integral formulation confirms the extended FIK identity's validity in capturing the influence of the coiled PSub vibrations on the dynamics of unstable channel flows, particularly through its interaction with flow perturbations.

\bibliography{RefPSubs}

\end{document}